\documentclass[prl,twocolumn,preprintnumbers]{revtex4-1}

\usepackage{epsfig}
\usepackage{graphicx}
\usepackage{color}
\usepackage[normalem]{ulem}
\usepackage[hidelinks]{hyperref}

\usepackage[caption=false]{subfig}

\begin{document}
\preprint{CERN-TH-2017-140}
\renewcommand{\thefigure}{\arabic{figure}}
\setcounter{figure}{0}

 \def\I{{\rm i}}
 \def\E{{\rm e}}
 \def\D{{\rm d}}

\bibliographystyle{apsrev}

\title{Bounding the speed of gravity with gravitational wave observations}

\author{Neil Cornish}
\affiliation{eXtreme Gravity Institute, Department of Physics, Montana State University, Bozeman, Montana 59717, USA}

\author{Diego Blas}
\affiliation{Theoretical Physics Department, CERN, CH-1211 Geneva 23, Switzerland}

\author{Germano Nardini}
\affiliation{AEC, Institute for Theoretical Physics, University of Bern, \\
Sidlerstrasse 5, CH-3012 Bern, Switzerland}

\begin{abstract}
The time delay between gravitational wave signals arriving at widely separated detectors can be used to place upper and lower bounds on the speed of gravitational wave propagation. 
Using a Bayesian approach that combines the first three gravitational wave detections reported by the LIGO Scientific \& Virgo Collaborations
we constrain the gravitational waves propagation speed $c_{\rm gw}$ to the 90\% credible interval $0.55 \, c  <  c_{\rm gw} <  1.42 \, c $, where $c$ is the speed of light in vacuum. These bounds will improve as more detections are made and as more detectors join the worldwide network. Of order twenty detections by the two LIGO detectors will constrain the speed of gravity to within 20\% of the speed of light, while just five detections by the LIGO-Virgo-Kagra network will constrain the speed of gravity to within 1\% of the speed of light.
\end{abstract}

\maketitle

The first detections of gravitational waves from merging black hole binaries~\cite{Abbott:2016blz,Abbott:2016nmj, Abbott:2017vtc} have been used to test many fundamental properties of gravity~\cite{Abbott:2017vtc, TheLIGOScientific:2016src,TheLIGOScientific:2016pea,Yunes:2016jcc}, and have been used to place the first observational upper limit on the speed of gravitational wave propagation~\cite{Blas:2016qmn}. In this letter we set a more stringent upper limit on the gravitational waves propagation speed $c_{\rm gw}$ by combining all the detections announced to-date, and by applying a full Bayesian analysis. We also provide the first direct {\em lower} bound on the propagation speed: $c_{\rm gw} > 0.55\, c$ at 95\% confidence.  While there are strong theoretical arguments that demand $c_{\rm gw} \geq c$ to prevent gravitational Cherenkov radiation~\cite{Moore:2001bv}, the LIGO detections provide the first direct observational constraints. 
%Models where the speed of gravitational waves differ from the speed of light generally predict a frequency dependence for $c_{\rm gw}$.

Gravitational waves generically propagate at a speed different from $c$ and with frequency dependence dispersion relations in theories of modified gravity, see e.g.~Refs.~\cite{Yunes:2016jcc,Blas:2016qmn,Ellis:2016rrr,Kostelecky:2016kfm,Bettoni:2016mij,deRham:2016nuf}. Thus, a precise determination of $c_{\rm gw}$ is a test of
gravitation complementary to other observations. To quantify what  `precise'  tests mean for  General Relativity, let us recall that some post-Newtonian parameters are known to
%\footnote{Certainly some modifications are better constrained, we just mention the order of magnitude of the worse constraints.}  
$\mathcal O(10^{-4})$ \cite{Will:2005va} while cosmological or other astrophysical observations typically constrain  modifications to General Relativity at the $\mathcal O(10^{-2})$ level \cite{Amendola:2016saw,Berti:2015itd}.

 A convenient parametrization for theories preserving rotation invariance is to write the dispersion relation as
\begin{equation}
\omega^2=m^2_g+c^2_{\rm gw}k^2+{a}\frac{k^4}{\Lambda^2}+...\,, 
\end{equation}
where $m_g$ refers to the mass of the graviton, $c_{\rm gw}$ is what we call `speed' of gravitational waves, and the rest of operators are wavenumber-dependent modifications suppressed by a high-energy scale $\Lambda$ (for a parametrization in scenarios breaking rotation invariance, see e.g.~Ref.~\cite{Kostelecky:2016kfm}).  Both $m_g$ and $\Lambda$ can be constrained by the absence of dispersion of the waves traveling cosmological distances. The scale $\Lambda$ is already constrained to be very large~\cite{Ellis:2016rrr}, making it very difficult to constrain the operator $a$. For the graviton mass the LIGO Scientific \& Virgo Collaborations put the strong bound $m_g < 7.7  \times 10^{-23}\, {\rm eV}/c^2$ \cite{Abbott:2017vtc}. However, the parameter $c_{\rm gw}$ can not be tested by dispersion measurements and other methods are required \cite{Blas:2016qmn}.

{\em Measuring $c_{\rm gw}$:}  In the following we focus on possible ways to directly measure $c_{\rm gw}$.
Since the signals measured by LIGO are dominated by the signal-to-noise accumulated in a narrow band between $50$ Hz -- $200$ Hz, our time delay bounds can be interpreted as constraints on the speed of gravity at a frequency $f\sim 100$ Hz. Since the LIGO bounds constrain dispersion effects to be small over hundreds of Mpc, they can safely be ignored on the terrestrial distance scales we are considering. Note that the inference that the observed signals come from hundreds of Mpc away relies on waveform models derived from General Relativity, and may not apply to a theory that
predicts $c_{\rm gw} \neq c$. 

The most obvious way to measure the speed of gravitational wave propagation is to observe the same astrophysical source using both gravity and light. However, for the three gravitational wave detections that have been announced thus far no unambiguous electromagnetic counterparts have been detected, and a different approach must be taken to constrain $c_{\rm gw}$. The finite distance between the Hanford and Livingston gravitational wave detectors can be used to set an absolute upper limit on the propagation speed~\cite{Blas:2016qmn} since the observed gravitational wave signals did not arrive simultaneously in the two detectors. Here we show that a proper statistical treatment that folds in the probability distribution of the time delays as function of $c_{\rm gw}$ also allows us to set lower bounds on the propagation velocity. It should be noted that when the first confirmed electromagnetic counterpart to a gravitational wave signal is finally observed, the bounds on the difference in propagation velocities, $|c_{\rm gw}-c|$ will be {\em many} orders of magnitude more stringent than what we can ever hope to set using gravitational wave signals alone~\cite{2014PhRvD..90d4048N,2016ApJ...827...75L,Ellis:2016rrr, Fan:2016swi}. Precisely for the same reason, this identification may never happen if the speed difference is not very small, since that would mean a significant time offset. For other possible stringent model-independent bounds not relying on the detection of a counterpart see \cite{Collett:2016dey,Bettoni:2016mij}.

{\em Constraints on $c_{\rm gw}$ from LIGO detections:}
The LIGO gravitational wave detectors at the Hanford and Livingston sites are separated by a light-travel time of $t_0 = 10.012 \, {\rm ms}$. The time delay for light along a propagation direction that makes an angle $\theta$ with the line connecting the two sites is $\Delta t_{\rm EM} =t_0 \cos\theta$. For sources distributed isotropically on the sky, there are equal numbers of sources per solid angle element $d\cos\theta\, d\phi$, thus the time delays for electromagnetic signals are uniformly distributed with $p(\Delta t_{\rm EM}) = 1/(2 t_0)$ for $-t_0 \leq \Delta t_{\rm EM} \leq  t_0$. The gravitational wave time delay is given by $\Delta t= (c/c_{\rm gw}) \Delta t_{\rm EM}$, thus the probability of observing a time delay $\Delta t$ between gravitational wave signals arriving at the two sites for sources uniformly distributed on the sky is given by the likelihood
\begin{equation}
p(\Delta t | c_{\rm gw}) =  \left \{\begin{array}{ll}
	 \frac{c_{\rm gw}}{2\, c \, t_0} \; &  {\rm for} \;  -\frac{ c t_0}{ c_{\rm gw}}  \leq \Delta t \leq  \frac{ c t_0}{ c_{\rm gw}}\, ,
	
	\\ &  \\
0 &    {\rm otherwise}. 
	\end{array} \right.
\end{equation}
While the sources may be uniformly distributed on the sky, the antenna patterns of the detectors make it more likely to detect systems above or below the plane of detectors. Assuming roughly equal sensitivity for the detectors, the observational bias scales as $F^3$, where $F(\theta,\phi)^2 = \sum_{D=H,L} F^D_{+,\times}(\theta,\phi)^2$ is the polarization averaged network antenna pattern~\cite{Schutz:2011tw}. The resulting distribution of electromagnetic time delays is then well fit by $p(\Delta t_{\rm EM}) = [1 - (\Delta t_{\rm EM}/t_q)^2][(3-t_0^2/t_q^2)\,2 t_0/3]^{-1}$ for $-t_0 \leq \Delta t_{\rm EM} \leq  t_0$ with $t_q=10.65$ ms. We use this modified distribution to define the likelihood $p(\Delta t | c_{\rm gw})$. For multiple events the full likelihood is the product of the per-event likelihoods. The posterior distribution for $c_{\rm gw}$ follows from Bayes' theorem: $p( c_{\rm gw}|\Delta t) = p(\Delta t | c_{\rm gw})  p(c_{\rm gw})/ p(\Delta t)$. We consider two possibilities for the prior on the speed of gravity, $p(c_{\rm gw})$: flat in $c_{\rm gw}$ and flat in  $\ln c_{\rm gw}$ in the interval $c_{\rm gw} \in [c_{\rm L},c_{\rm U}]$. For the results shown here we set $c_{\rm U} = 100 \, c$, and either $c_{\rm L} = c/100$, or $c_{\rm L} = c$. The latter limit takes into account the Cherenkov radiation constraint~\cite{Moore:2001bv}. For three or more events the choice of prior has very little impact on the upper limit. To account for the measurement error in $\Delta t$ we use a Markov Chain Monte Carlo to marginalize over the errors in the arrival times.
\begin{figure}[htp]
\includegraphics[clip=true,angle=0,width=0.47\textwidth]{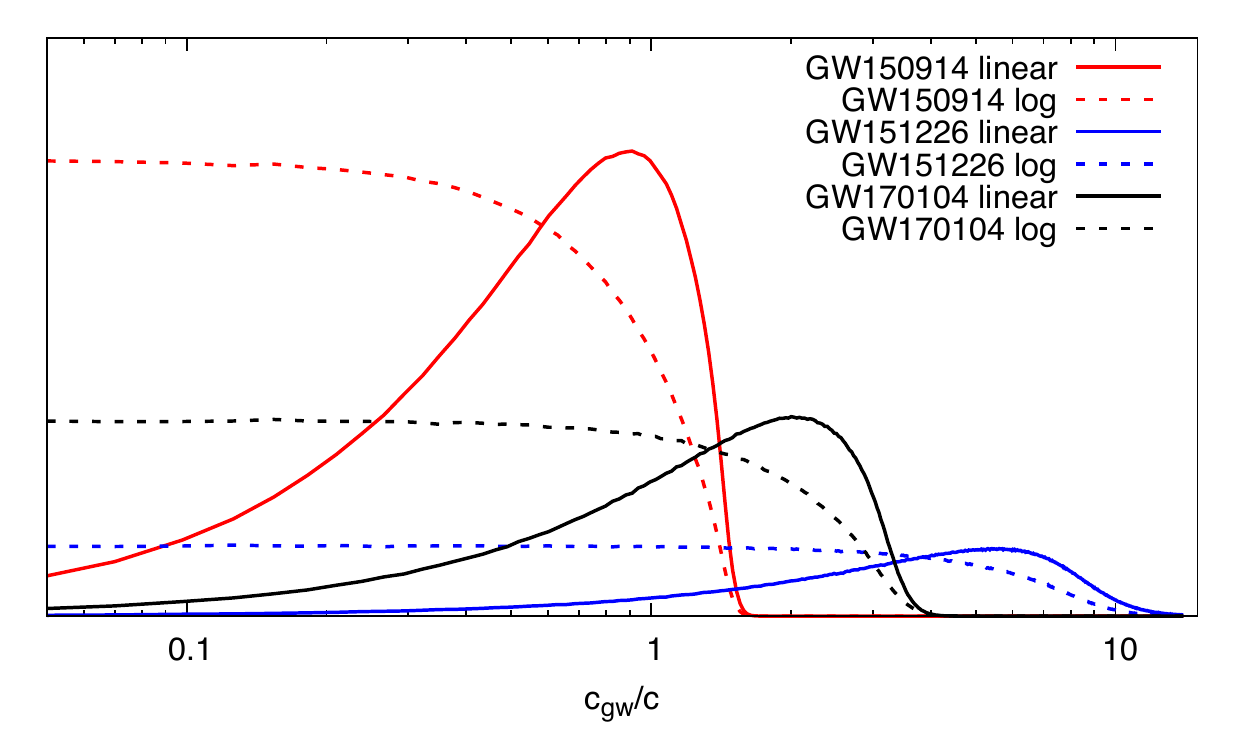} 
\caption{\label{fig:ind} Posterior distributions for the gravitational wave propagation speed derived from each of the individual LIGO events for prior distributions uniform in $c_{\rm gw}$ or $\ln c_{\rm gw}$.}
\end{figure}

The first detections of black hole mergers by LIGO provide measurements of $\Delta t$ that were quoted in terms of central values and 90\% credible intervals. Since the full posterior distributions for
$\Delta t$ were not provided, we assume that the distributions can be approximated as normal distributions with mean $\mu$ and standard deviation $\sigma$ with values:
GW150914 ($\mu = 6.9\,{\rm ms}$, $\sigma= 0.30\, {\rm ms}$)~\cite{TheLIGOScientific:2016pea}; GW151226 ($\mu = 1.1\,{\rm ms}$, $\sigma= 0.18\, {\rm ms}$)~\cite{TheLIGOScientific:2016pea}; GW170104 ($\mu = 3.0\,{\rm ms}$, $\sigma=  0.30\, {\rm ms}$)~\cite{Abbott:2017vtc} (for a discussion about this assumption, see e.g.~Ref.~\cite{Cutler:1994ys}). The upper bound on $c_{\rm gw}$ quoted in Ref.~\cite{Blas:2016qmn} was found by taking the minimum time delay from GW150914 as $\Delta t = \mu - 2\sigma = 6.3$\,ms, and demanding that $c_{\rm gw} < c\, t_0/\Delta t =1.6\,c$. Note that this value is lower than the bound of $1.7\, c$ quoted in Ref.~\cite{Blas:2016qmn} as they interpreted the error in $\Delta t$ quoted in Ref.~\cite{Abbott:2016blz} as one-sigma errors, when in fact they were the bounds on the 90\% credible interval.

\begin{figure}[htp]
\includegraphics[clip=true,angle=0,width=0.46\textwidth]{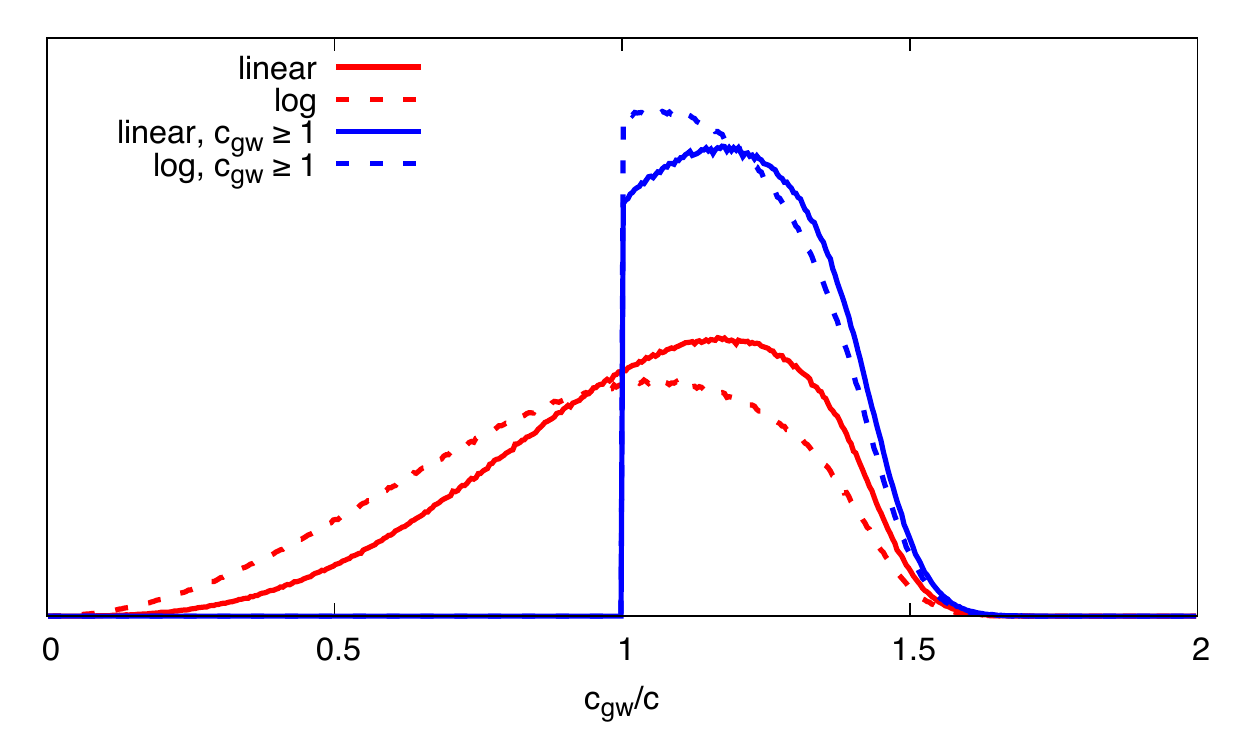} 
\caption{\label{fig:combo} Posterior distributions for the gravitational wave propagation speed derived by combining the first three LIGO detections. Prior distributions uniform in $c_{\rm gw}$ or uniform in $\ln c_{\rm gw}$ were considered, with the interval starting at either $c_{\rm L} = c/100$ or $c_{\rm L} = c$. }
\end{figure}

We compute the posterior distribution for the gravitational wave propagation speed, $p(c_{\rm gw} | \Delta t )$ using a Markov Chain Monte Carlo algorithm that marginalizes over the uncertainties in the time delays by drawing new values of $\Delta t$ from the assumed posterior distributions at each iteration of the Markov chain. Figure \ref{fig:ind} shows the posterior distributions for $c_{\rm gw}$ using each of the detections separately.
Individually the three events yield 95\% upper bounds on the propagation velocity for the linear and (log) uniform priors of $c_{\rm gw} < 1.37\, c\; (1.26\, c)$ for GW150914;
$c_{\rm gw} < 10.1\, c \; (8.57 \, c)$ for GW151226; and $c_{\rm gw} < 3.19\, c\; (2.94\, c)$ for GW170104.  Each event also yields a 95\% lower bound on the propagation velocity, but these limits are not very interesting since all of the distributions have some support at $c_{\rm gw}  \simeq 0$. Note that GW151226 produces the weakest upper bound on the propagation velocity even though it has the most accurately measured time delay. This is because the strongest upper limits come from events with the longest time delay, and even allowing for the uncertainties in the time delay measurements for GW150914 and GW170104, both are constrained to have delays that are much longer than for GW151226.

Figure \ref{fig:combo} shows the posterior distribution for $c_{\rm gw}$ found by combining all three LIGO detections together for uniform priors in $c_{\rm gw}$ or $\ln c_{\rm gw}$. For the wider prior range with $c_{\rm L} = c/100$ the combination of the three detections yield an interesting lower bound on the propagation speed. The 90\% credible interval for the linear and log priors are $0.55 \, c  <  c_{\rm gw} <  1.42 \, c $ and $0.41 c  <  c_{\rm gw} <  1.39 \, c $ respectively. The upper limit is only weakly dependent on the choice of prior distribution. Notice that the form of the posterior distributions can be qualitatively understood analytically by using Bayes'  theorem and ignoring the complications from antenna patterns and noise. 
For a uniform prior on $c_{\rm gw}$ this gives $p(c_{\rm gw} | \Delta t_1, \Delta t_2, \Delta t_3) = 4 c_{\rm gw}^3 / c_*^4$ for $ c_{\rm gw} < c_*$, where $c_* = c t_0/\Delta t_1$ and $\Delta t_1$ is the longest of the observed time delays. For the uniform-in-log prior the posterior distribution is $p(c_{\rm gw} | \Delta t_1, \Delta t_2, \Delta t_3) = 3 c_{\rm gw}^2 / c_*^3$ for $ c_{\rm gw} < c_*$. This explains the growing of the posterior in Figure~\ref{fig:combo}, while the smoothing of the curve is related to the Gaussian noise. While an analytical understanding of the posterior distribution is possible, it is necessary to use numerical sampling methods to compute the {\it full} shape that includes marginalization over observational errors, possible orientation biases, and add new detectors and detections.

One possible limitation of using the published LIGO results, rather than analyzing the raw data, is that the standard LIGO searches exclude signals with Hanford-Livingston time delays greater 15 ms~\cite{TheLIGOScientific:2016qqj}, so potentially missing some signals if $c_{\rm gw} <  0.66 c$. On the other hand, loud single detector triggers consistent with a binary merger would not
go un-noticed, as evidence by GW170104, which was missed by the standard search due to the Hanford detector being incorrectly flagged as out of observing mode, and only found later in an analysis of single detector triggers from the Livingston detector~\cite{Abbott:2017vtc}. It is highly unlikely that pairs of triggers with time delays greater than 15 ms would be overlooked, especially if they shared similar parameters and occurred within minutes of each other.

{\em Forecasts for more detections and more detectors:} 
It is interesting to consider how the bounds will improve with additional detections and detectors, using just the gravitational time delays (as mentioned earlier, combined electromagnetic and gravitational observations will dramatically improve the measurements). The upper bound is mostly set by detections with large time delays, while the lower bound is set by having many signals with a wide range of delays. For the two-detector LIGO network, and assuming $c_{\rm gw}=c$ as predicted by General Relativity, we should see one event with a time delay $\Delta t > 8.6\,$ms with ten detections, and one event with a time delay $\Delta t > 9.6\,$ms with 100 detections. Just those single events would yield 99\% upper limits better than $c_{\rm gw} <  1.2 \, c $ and $c_{\rm gw} <  1.07 \, c $ respectively, assuming that $\Delta t$ is measured to a level of accuracy typical of the first three detections (the accuracy with which $\Delta t$ can be measured depends on the signal-to-noise ratio and the number of cycles completed in-band, among other things). Performing multiple Monte-Carlo simulations under the assumption that General Relativity is the correct theory of gravity indicates that with 100 detections by the 2-detector LIGO network we will be able to constrain $c_{\rm gw}$ to within a few percent of the speed of light for both the upper and lower bounds.

\begin{figure}[htp]
\includegraphics[clip=true,angle=0,width=0.46\textwidth]{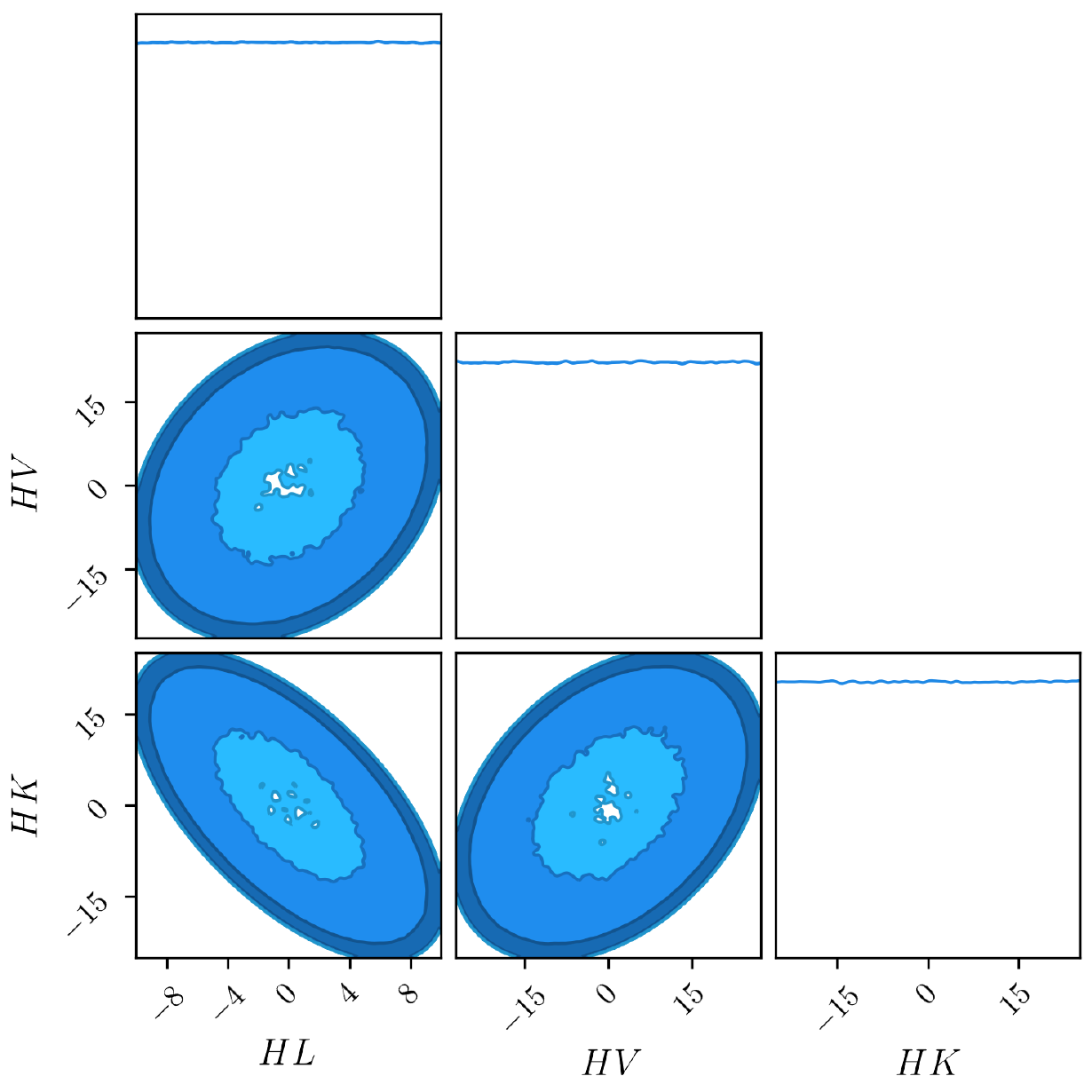} 
\caption{\label{fig:delays} Slices through the joint electromagnetic time delay distribution for the HLVK network using Hanford as the reference site. Darker colors in the two-dimensional slices indicate higher density. Note the cup-like structure of the distributions (higher at the edges than in the center).}
\end{figure}

Far better constraints can be achieved with far fewer events by using a larger network of detectors. In the next few years the LIGO Hanford (H) and Livingston (L) detectors will be joined by the Virgo (V) detector in Italy, the Kagra (K) detector in Japan~\cite{KAGRA:2016rjn}, and later, another LIGO detector in India~\cite{Aasi:2013wya}. With an $N$ detector network there are $N(N-1)/2$ time delays. From the geopositions of the detector sites \cite{geolocalization} the maximum electromagnetic time delays between these sites are: $HL = 10.012 \, {\rm ms}$, $HV = 27.288\, {\rm ms}$, $HK =  25.158\, {\rm ms}$, $LV = 26.448 \, {\rm ms}$, $LK = 32.455 \, {\rm ms}$, $VK = 29.202 \, {\rm ms}$. (LIGO India would further improve the bound, but will not be available at least 2025).
Upper bounds on the speed of gravity are dominated by events with sky locations that come close to maximizing the electromagnetic time delay between a pair of detectors. For the HL network just 5\% of events are within 95\% of the maximum time delay, while for the HLVK network 25\% of events are. Thus, on average, it only takes a few events to produce tight limits using the larger network of detectors.
 Complete information about the inter-site time delays is contained in the joint probability distribution of the $N-1$ electromagnetic  time delays between one reference detector and the other detectors in the network. Figure \ref{fig:delays} shows slices through the joint electromagnetic time delay distribution for the HLVK network using Hanford as the reference site assuming a uniform distribution of sources. Here we did not correct for the observational bias, since the network antenna pattern for a four detector network is fairly uniform. Applying the change of variable $\Delta t= (c/c_{\rm gw}) \Delta t_{\rm EM}$ to this distribution as we did for the two-detector HL case yields the joint likelihood $p(\Delta t_{\rm HL}, \Delta t_{\rm HV}, \Delta t_{\rm HK} | c_{\rm gw})$. Using simulated detections of events measured to a precision of $\sigma = 0.3\, {\rm ms}$ in each detector, we find that with just 3 detections the HLVK network will typically be able to constrain $c_{\rm gw}$  to the 99\% credible region $c_{\rm gw}/c = 1.00\pm 0.02$. The constraints improve to better than 1\% of the speed of light with 5 detections. 
 
{\em Summary:} Combining the time delay measurements between detector sites for multiple gravitational wave events can be used to place interesting constraints on the speed of gravity. The LIGO detections made to-date already constrain the speed of gravity to within 50\% of the speed of light. Additional LIGO detections in the next few years should improve the bound to of order 10\%. The bounds will improve rapidly as more detectors join the worldwide network, with just a half-dozen detections by the Hanford-Livingston-Virgo-Kagra network constraining deviations to better than 1\%. These bounds will allow to test General Relativity to the level of other standard tests, as those coming from the damping of orbits in binary systems or cosmology.

\section*{Acknowledgments}
The authors would like to thank the Centro de Ciencias de Benasque Pedro Pascual for providing a wonderful venue to carry out this work. The authors appreciate the constructive feedback received from Walter Del Pozzo, Thomas Dent, John Veitch, and Brian O'Reilly during the internal LIGO review. DB is grateful to  V\'ictor Planas-Bielsa and Sergey Sibiryakov for discussion. NJC appreciates the support of NSF award PHY-1306702. GN is supported by the Swiss National Science Foundation (SNF) under grant 200020-168988. The authors thank the LIGO Scientific Collaboration for access to the data and gratefully acknowledge the support of the United States National Science Foundation (NSF) for the construction and operation of the LIGO Laboratory and Advanced LIGO as well as the Science and Technology Facilities Council (STFC) of the United Kingdom, and the Max-Planck-Society (MPS) for support of the construction of Advanced LIGO. Additional support for Advanced LIGO was provided by the Australian Research Council.

\bibliography{refs}

\end{document}